\documentclass{emulateapj}

\newcommand{\tsc}{\textsc}
\newcommand{\rhalf}{$r_{1/2}$\,}
\newcommand{\Rhalf}{${\mathrm{R}}_{1/2}$\,}
\newcommand{\rhalfstar}{$r^{*}_{1/2}$\,}
\newcommand{\Rc}{R\,}
\newcommand{\mtot}{M$_{\mathrm{tot}}$\,}
\newcommand{\mtotL}{M$_{\mathrm{tot}}$/L\,\,}
\newcommand{\rhalfRc}{($r_{1/2}, {\rm{R}}$)\,}
\newcommand{\RhalfRc}{(${\mathrm{R}}_{1/2}, {\rm{R}}$)\,}
\newcommand{\coma}{\textsc{coma}\,}
\newcommand{\virgo}{\textsc{virgo}\,}

\newcommand{\snap}{\textsc{snap}}

\begin{document}
\bibliographystyle{apj}

\title{Probing the truncation of galaxy dark matter halos in high density environments 
from hydrodynamical N-body simulations}

\author{
Marceau Limousin \altaffilmark{1},
Jesper Sommer-Larsen \altaffilmark{1},
Priyamvada Natarajan \altaffilmark{2,3} \&
Bo Milvang-Jensen \altaffilmark{1}
}

\altaffiltext{1}{Dark Cosmology Centre, Niels Bohr Institute,
University of Copenhagen, Juliane Maries Vej 30, 2100 Copenhagen,
Denmark; marceau@dark-cosmology.dk} \altaffiltext{2}{Astronomy
Department, Yale University, P.O. Box 208101, New Haven, CT
06520-8101, USA} \altaffiltext{3}{Department of Physics, Yale
University, P.O. Box 208101, New Haven, CT 06520-8101, USA}

\begin{abstract}
We analyze high resolution, N-body hydrodynamical simulations of
fiducial galaxy clusters to probe tidal stripping of the dark matter
subhalos. These simulations include a prescription for star formation
allowing us to track the fate of the stellar component as well. We
investigate the effect of tidal stripping on cluster galaxies hosted
in these dark matter subhalos as a function of projected cluster-centric radius.
To quantify the extent of the dark matter halos of cluster galaxies, we introduce the
half mass radius \rhalf as a diagnostic, and study its evolution with
projected cluster-centric distance \Rc as a function of redshift. We
find a well defined trend for \rhalfRc : the closer the galaxies are
to the center of the cluster, the smaller the half mass
radius. Interestingly, this trend is inferred in \emph{all} redshift
frames examined in this work ranging from $z$\,=\,0 to $z$\,=\,0.7.
At $z$\,=\,0, galaxy halos in the central regions of clusters are
found to be highly truncated, with the most compact half mass radius
of 10\,kpc. We also find that \rhalf depends on luminosity and we
present scaling relations of \rhalf with galaxy luminosity. The
corresponding total mass of the cluster galaxies is also found to
increase with projected cluster-centric distance and luminosity, but
with more scatter than the \rhalfRc trend. Comparing the distribution
of stellar mass to total mass for cluster galaxies, we find that the
dark matter component is preferentially stripped, whereas the stellar
component is much less affected by tidal forces. 
We compare these results with galaxy-galaxy lensing probes
of \rhalf and find qualitative agreement. Future surveys with space
based telescopes such as \textsc{dune} and \snap, that combine wide
field and high resolution imaging, will be able to probe the predicted
\rhalfRc relation observationally.
\end{abstract}

\section{Introduction}

The dependence of galaxy properties on environment is well established
\citep[see, \emph{e.g.}][]{adami98,lanzoni,bosellireview,delucia}.
One of the most extreme environments for galaxies is inside a massive
galaxy cluster, where active, strong tidal forces are exerted by the
global cluster potential. The theoretical expectation is that the global
tidal field of a massive, dense cluster potential is strong
enough to truncate the dark matter halos of galaxies that traverse the
cluster core. Early work
\citep[\emph{e.g.}][]{merritt83,richstone84,merritt84} found that
a large fraction of the mass initially attached to galaxies in the
central Mpc is stripped.
\citet{avila99,bullock,avila05} found that halos in dense environments
are more truncated and more compact than their isolated counterparts
of the same luminosity. On galaxy scales, using dark matter only
simulations, \citet{diemand07} have studied the evolution of subhalos
in the Via Lactea host halo, and find that tidal forces remove subhalo
mass from the outside in, which leads to higher concentrations for
subhalos located in the inner regions compared to field halos of the
same mass. Detailed studies of the evolution of subhalos
in clusters in dark matter only simulations have been performed
by \citet[][hereafter G98]{ghigna98} and \citet{DeLucia04}.  
G98 find that the dominant interactions in cluster environments 
are between the global cluster tidal field and individual galaxies after
$z$\,=\,2, and that the cluster tidal field significantly strips galaxy
halos. Moreover, both numerical simulations (G98) and
analytical calculations \citep{mamon00} predict that the tidal radius
of a given galaxy depends on its cluster-centric distance. As a
consequence, the closer to the center the galaxies are, the stronger
are the tidal forces they will experience, resulting in more compact
subhalos. G98 probed the characteristic extent of
galaxy subhalos with cluster-centric distance, at $z$\,=\,0
and $z$\,=\,0.5. Considering the three
dimensional cluster-centric distance, they find that subhalo extents
decreases towards the cluster center, but this trend was weak and
therefore hard to detect at $z$\,=\,0.5. They argue that this is due to
the fact that at $z$\,=\,0.5, the cluster has quite an anisotropic mass
distribution and tides are efficient only at its very center. When
considering the \emph{projected} cluster-centric distance, this trend
does not totally disappear but becomes marginal, suggesting that
subhalo extents do not strongly depend on present day projected
cluster-centric distances. 

From an observational point of view, gravitational lensing seems the
only currently viable method to probe the extent of dark matter
subhalos in clusters, and this statement is specially true for cluster
galaxies. In
the case of field galaxies, satellite dynamics can be used as well
\citep[\emph{e.g.}][]{prada} and gives results consistent with current
galaxy-galaxy lensing studies. The deflection caused by galaxy scale
mass concentrations is small (quantified by a shear $\gamma \sim
0.01$) and thus is challenging to detect. However, galaxy-galaxy
lensing studies in clusters have been performed
\citep{Priya1,geigeramas,Priya2,Priya3,mypaperII} where 
this shear signal is boosted by the smoothly distributed large scale dark
matter distribution.
These studies have successfully statistically
detected the weak lensing signal generated by cluster galaxies
\citep[see also recent work by][]{aleksi}.
These analyses have provided evidence for truncation of
galaxy dark matter halos in high density environments. The inferred
typical half mass radius was found to be typically more compact than 50\,kpc, 
whereas half mass radii larger than 200\,kpc are derived for field galaxies of
equivalent luminosity \citep{fischer,mckay,hoekstra03,hoekstra04}. 
Moreover, when using the \textsc{nfw} \citep{nfw} profile, concentration 
parameters greater than 20 were inferred \citep{moriond} for some clusters.
These results are in qualitative agreement with the tidal striping
scenario. Recently, using a large \tsc{hst} mosaic covering up to
5\,Mpc from the center of galaxy cluster Cl0024 at $z$\,=\,0.39, Natarajan
et~al. (2008) were able to probe the galaxy population in three radial
bins and inferred a larger extent for the halos of galaxies living in
the outskirts of the cluster (i.e. at a cluster-centric distance
between 3 and 5 Mpc) compared to the galaxies living in the core of the 
cluster (here between 0 and 3 Mpc).

In this paper, we want to investigate how tidal forces shape the fate
of galaxy dark matter halos, and in particular, we focus on the
\emph{extent} of cluster galaxy dark matter halos and probe its
evolution with cluster-centric distance and redshift. We analyze
N-body hydrodynamical simulations of two fiducial galaxy
clusters that contain not only dark matter particles but also stars
and gas particles that interact through many physical processes. In
order to make predictions for future surveys that will constrain the
extent of the halos as a function of cluster-centric distance, we
consider the \emph{projected} cluster-centric distances since this is
what can be probed observationally.  The extent of subhalos is
quantified by the half mass radius \rhalf since it is directly
comparable to the characteristic extent probed via galaxy-galaxy
lensing studies. 

This paper is organized as follows: in Section 2, we discuss the
properties of the two simulated clusters studied here. In Section 3,
we investigate how the half mass radius \rhalf and the total mass
\mtot evolve with cluster-centric distance, for different assembly
stages (redshift frames) in these simulations. In Section 4, we
investigate how \rhalf and \mtot evolve with luminosity. We fit
scaling relations for \rhalf, \mtot and the luminosity, and present
total mass to light ratios. In
Section 5, we investigate the ratio of the total to the stellar mass
and its evolution with cluster-centric distance and redshift. In
Section 6, we compare galaxy-galaxy lensing results to our theoretical
predictions, and finally we present our conclusions in Section 7. All
our results in this paper are scaled to the flat, low matter density
$\Lambda$\tsc{cdm} cosmology with $\Omega_{\mathrm{M}} = 0.3, \, \Omega_\Lambda =
0.7$ and a Hubble constant ${\mathrm{H}}_0 = 70$ km\,s$^{-1}$ Mpc$^{-1}$.

\section{Simulations of our fiducial clusters}

\subsection{Cluster Properties}
N-body hydrodynamical (\tsc{treesph}) simulations of the
formation and evolution of two galaxy clusters in a $\Lambda$\tsc{cdm} cosmology
have been performed. The simulations include metallicity-dependent
radiative cooling, star formation according to different initial mass
functions, energy feedback as strong starburst-driven galactic
superwinds, chemical evolution with non-instantaneous recycling of gas
and heavy elements, effects of a metagalactic ultraviolet field and
thermal conduction in the intracluster medium. For full details on
these simulations and for a comparison of the properties of the
simulated clusters to observations, we refer the reader to a series
of three papers: \citet{romeo1,romeo2,romeo3}, hereafter \textsc{rsl}.

The main difference between the two simulated clusters is their
temperature (or equivalently their mass) and the numerical
resolution. The more massive cluster is a 6\,keV cluster, that we will
hereafter refer to as \tsc{coma}. At $z$\,=\,0, its virial radius is
equal to 2.9\,Mpc, and the corresponding virial mass is equal to
1.3\,$\times$\,10$^{15}$\,M$_{\sun}$. The second cluster is a 3.1\,keV
cluster that we will refer to as \tsc{virgo}.  At $z$\,=0\,, its
virial radius is equal to 1.8\,Mpc, and the corresponding virial mass
is equal to 2.8\,$\times$\,10$^{14}$\,M$_{\sun}$. The particle mass in
\virgo is 8 times lower than the one used to simulate \coma.  Thus the
\virgo simulation is a higher resolution one and we are able to
resolve smaller galaxies.  On the other hand, since \coma is a more
massive cluster, more massive galaxies than in \virgo will form.

The highest redshift frame we analyze is $z$\,=\,0.7. At this
redshift, both clusters are virialized and there is a clear dominant
cluster halo. At higher redshifts, the clusters are not well defined
(no clear dominant halo) and are still under construction with major
merger events. However, for any redshift slice, \coma is less relaxed
than \virgo. In the case of \coma at $z$\,=\,0.7, simulations reveal
some infalling galaxies, whereas that is not the case in \virgo. This
difference in dynamical states can be quantified by the redshift of
the last major merger event: for \coma, the last merger event occurs
at around $z$\,=\,0.8, whereas it is at around $z$\,=\,1.5 for
\virgo. This means that \coma is dynamically younger than \virgo, and,
as we will see, the \virgo galaxy population is more relaxed, with
less scatter in the relations we study in this work.  Due to the
difference in dynamical states between the two simulated clusters, it
is not easy to directly compare their respective properties, in the
sense that \coma is not simply a rescaling of \virgo at a higher
temperature.
Pictures and movies of the simulated clusters can be found at
http://www.tac.dk/~jslarsen/Clusters/

\subsection{Extracting subhalos from the simulated clusters}

The first step of our analysis is to extract substructures from the
simulated cluster in order to get a catalog of cluster galaxies. This
procedure is detailed in \textsc{rsl} and summarized below.
In order to ascertain that we identify all galaxies, and hence sub-halos,
in the main cluster halo the following approach was adopted. A cubic grid
of cube-length $\Delta l$=10 kpc is overlaid the cluster, and all cubes
containing at least $N_{\rm{th}}$=2 star particles are identified.
Subsequently, each selected cube is embedded in a larger cube of cube-length 3$\Delta l$.
If this larger cube contains at least $N_{\rm{min}}$=7 star particles, which are
gravitationally bound by its content of gas, stars and dark matter the system is
identified as a potential galaxy. Since the method can return several, almost identical
versions of the same galaxy only the one containing the largest number of star particles
is kept and classified as a galaxy. We tested the galaxy identification algorithm by
varying $\Delta l$, $N_{\rm{th}}$, $N_{\rm{min}}$, and also the numerical resolution,
and found it to be adequately robust for the purposes of this paper. 

\subsection{Subhalos Properties}
Once the subhalos have been
identified, galaxy half mass radii, \rhalf, and total masses, \mtot,
were determined by an iterative approach, described below. For a given
galaxy the density of the smooth cluster "background" of dark matter
and hot, intra-cluster gas at the position of the galaxy is
determined. This is done by calculating the average total density in 5
kpc thick shells of radii from $r_s$ to 1.5\,$r_s$. The "search
radius", $r_s$, is initially set to 4\,$r_t$, where $r_t$, is the
tidal radius of the galaxy at its position in the cluster defined by \citep{binney}:
\begin{equation}
r_t \,\, =  r_p \,\, \left( \frac{m}{3\,M} \right)^{1/3}
\label{rtequ1}
\end{equation}

All shells containing other substructures, i.e., galaxies and/or dark
matter halos, are subsequently removed using a background density
fluctuation criterion, and finally the background density is
determined from the remaining shells.  Second, the cumulative mass of
the galaxy, including its dark matter halo (above the background
density) is determined in 2.5 kpc thick shells, going from inside and
out, until a shell is reached of density equal to or less than that of
the background. Using the cumulative mass distribution, the half mass
radius \rhalf, is subsequently determined by linear interpolation.
Third, the search radius is set to 4\,\rhalf, and the above procedure
is redone. This is done repeatedly until convergence. Tests with
different parameter values show the above procedure to be an efficient and 
robust way for determining half mass radii and total masses of cluster galaxies.

By construction, \rhalf is the radius within which the 3D mass of the
simulated galaxies equals half of the total mass. We will refer to
this quantity as being the 3D half mass radius. We also pursue a 2D
analysis. In particular, we
looked at \Rhalf, the radius within which the 2D mass equals half of
the total mass, as described below. Once we know the 3D region
(assumed spherical) where a given galaxy is confined and contained, we
projected the excess mass density in this region (above the cluster
mean density) onto the plane along the three cardinal directions, and
averaged over the results. This gives us a robust estimate of the 2D
projected mass, that we use to calculate \Rhalf. We will refer to
this quantity as the 2D half mass radius.  We will compare these
quantities in Section 3. However, when looking at possible scalings of
the half mass radius with luminosity, we will use \rhalf only. The
quantity \rhalf is robust as it is easily calculated and well defined
for most popularly used mass models. Moreover, \rhalf is not the result 
of an average of projected quantities, thus in principle a less noisy 
quantity than \Rhalf.

There is a large scatter in the properties of the galaxies that form
in these cluster simulations. This scatter is mainly due to the
diverse orbital histories of galaxies.  To consolidate results, we
define bins, in which we compute the median of the points as well as
the standard deviation $\sigma$. The error is defined as
$\sigma/\sqrt{\textsc{n}}$, where \textsc{n} is the number of objects
in a given bin.  In the following, when needed, we fit a linear
function to the points corresponding to the galaxies, and not to the
binned data.

\subsection{Projected cluster-centric distance}
\begin{figure}[h!]
\epsscale{1.0}
\plotone{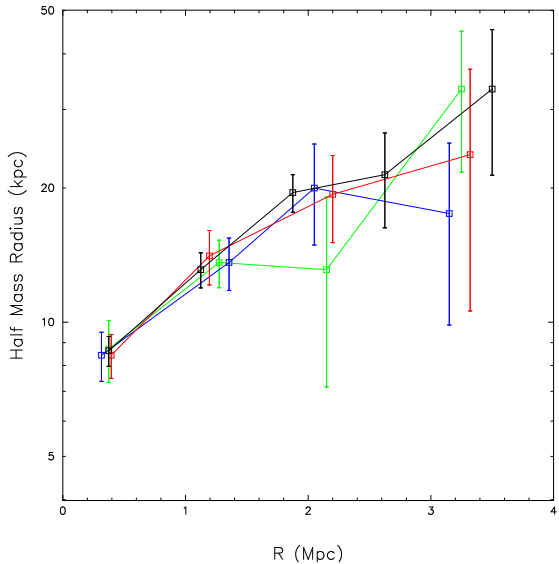}
\caption{
median half-mass radii \rhalf as a function of projected cluster-centric distance
in \virgo.
This cluster-centric distance corresponds to the projection of the 3D cluster-centric
distance along the three cardinal axes: along the $x$ axis (green), the $y$ axis (red) and
the $z$ axis (blue).
Also show is the relation when merging the three data sets (black) which corresponds
to the solid line in the upper left panel of Fig.~2. We see no bias with viewing angle.
}
\label{3projections}
\end{figure}

Since observational probes are sensitive to \emph{projected}
cluster-centric distances, we project the 3D radial distance of
the galaxies with respect to the center of the cluster over the three
cardinal axes, and our cluster galaxy catalog is made by merging the
three catalogs corresponding to each projection. We thus artificially
increase the data by a factor of three. This is not unreasonable to do
since we can imagine that three different observers have been looking
at the cluster from the three cardinal axes and have merged their
corresponding data sets. This procedure is equivalent to observing
three different clusters in a similar dynamical state.
It is important at this point to make sure that the relation between
half mass radius and projected cluster-centric distance we aim to probe
does not depend on the adopted viewing angle and that we do not
introduce any systematics by merging catalogs corresponding to each
projection.
To this purpose, we have looked at the evolution of the half mass
radius with respect to the three different projections of the cluster-centric
distance. We find no bias. To illustrate this point, we show on
Fig.~\ref{3projections} the evolution of the half mass radius as a function
of the projected cluster-centric distances, considering the binned data and the
\virgo cluster.
We see that the relation does not depend on the adopted projected cluster-centric
distance. The same conclusion can be drawn when considering the unbinned data points.
Moreover, we reach similar conclusions in the case of the \coma cluster.
Also shown for comparison is the relation when merging the
data sets as done in the rest of the paper.
Note that looking at possible variations with viewing angle is interesting
from an observational perspective and suggests that projection effects
are not important for observational studies.

\section{Trends with cluster-centric distance}

\begin{figure*}[h!]
\epsscale{0.5}
\plotone{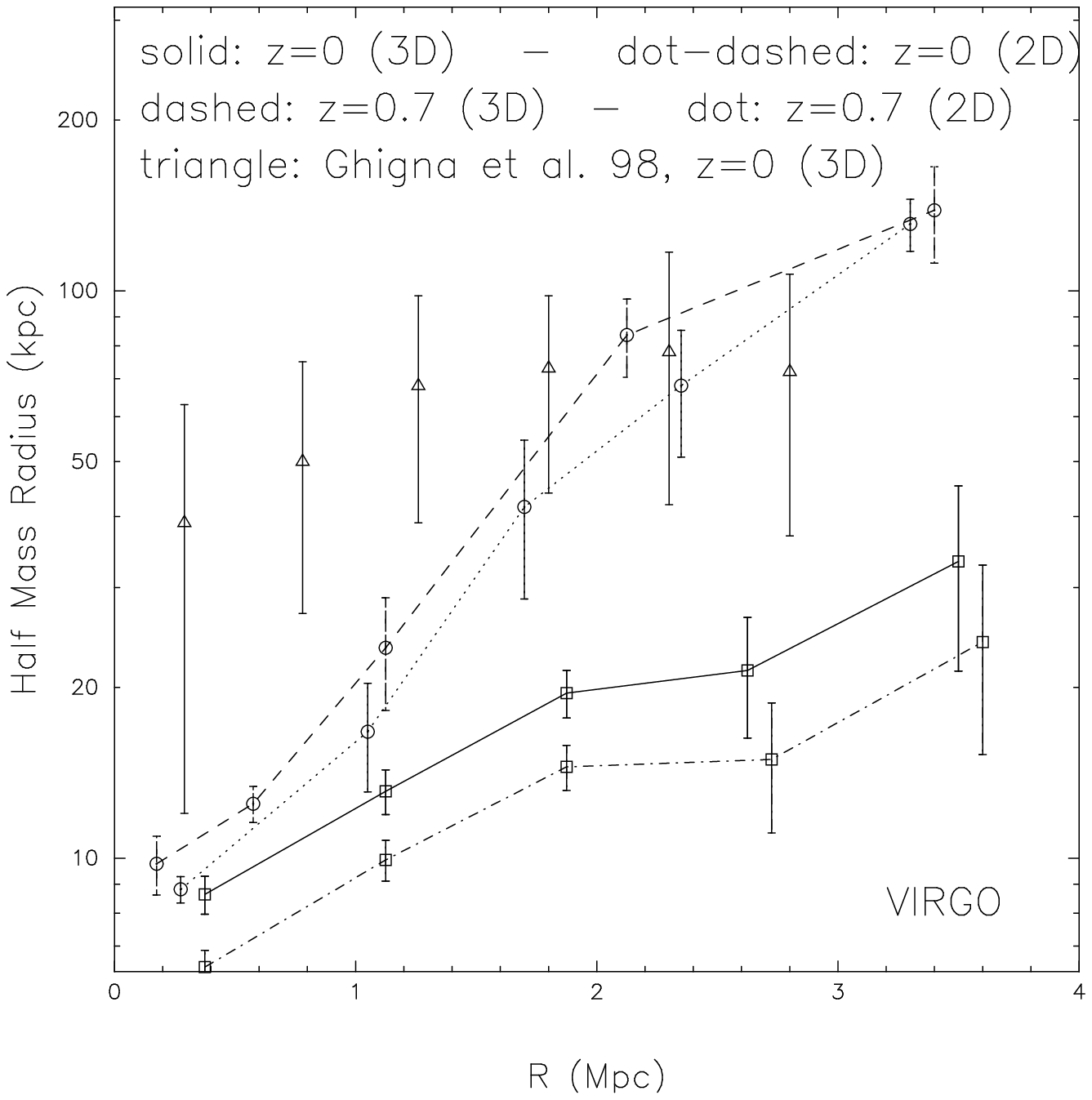}
\plotone{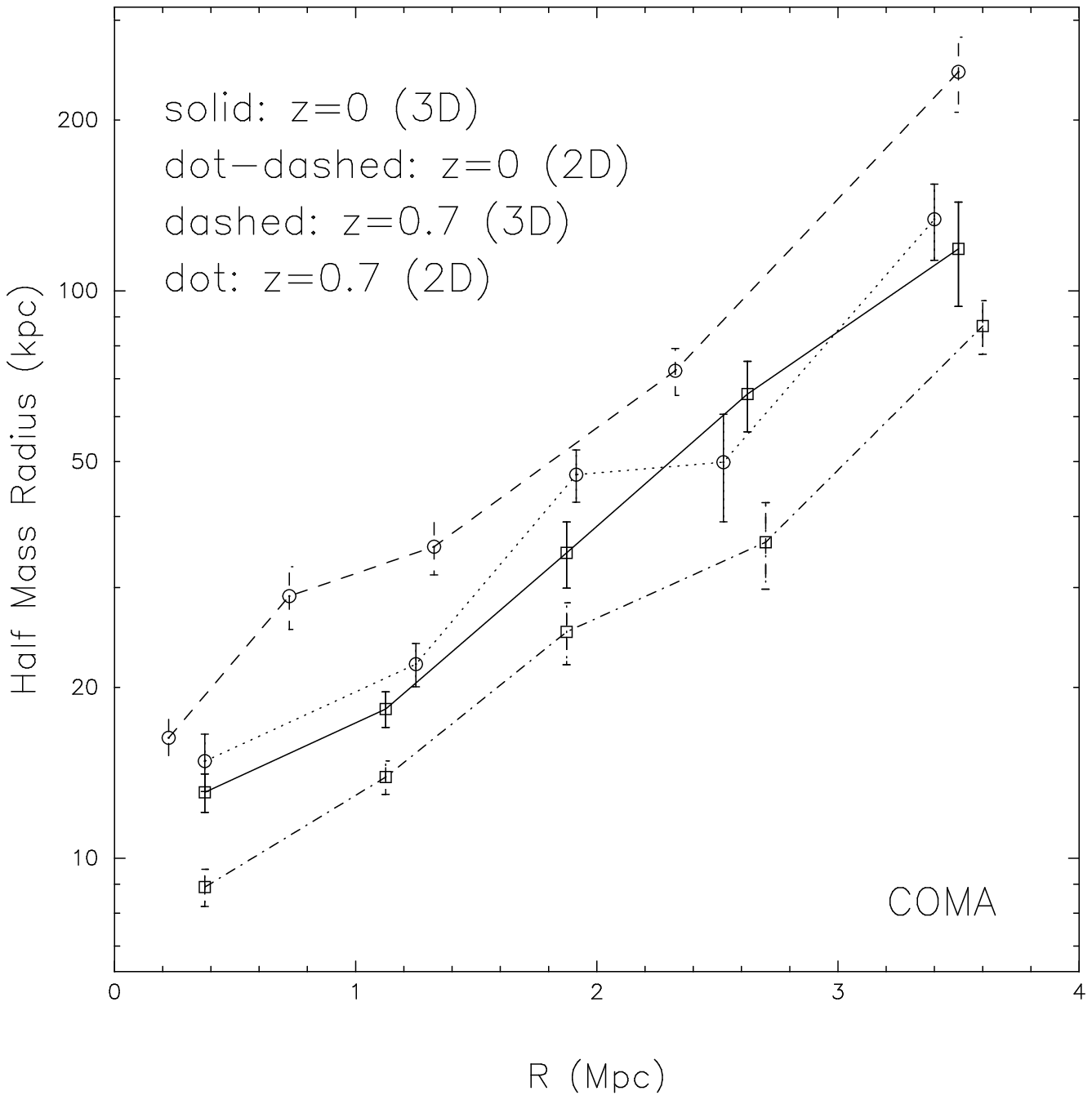}
\plotone{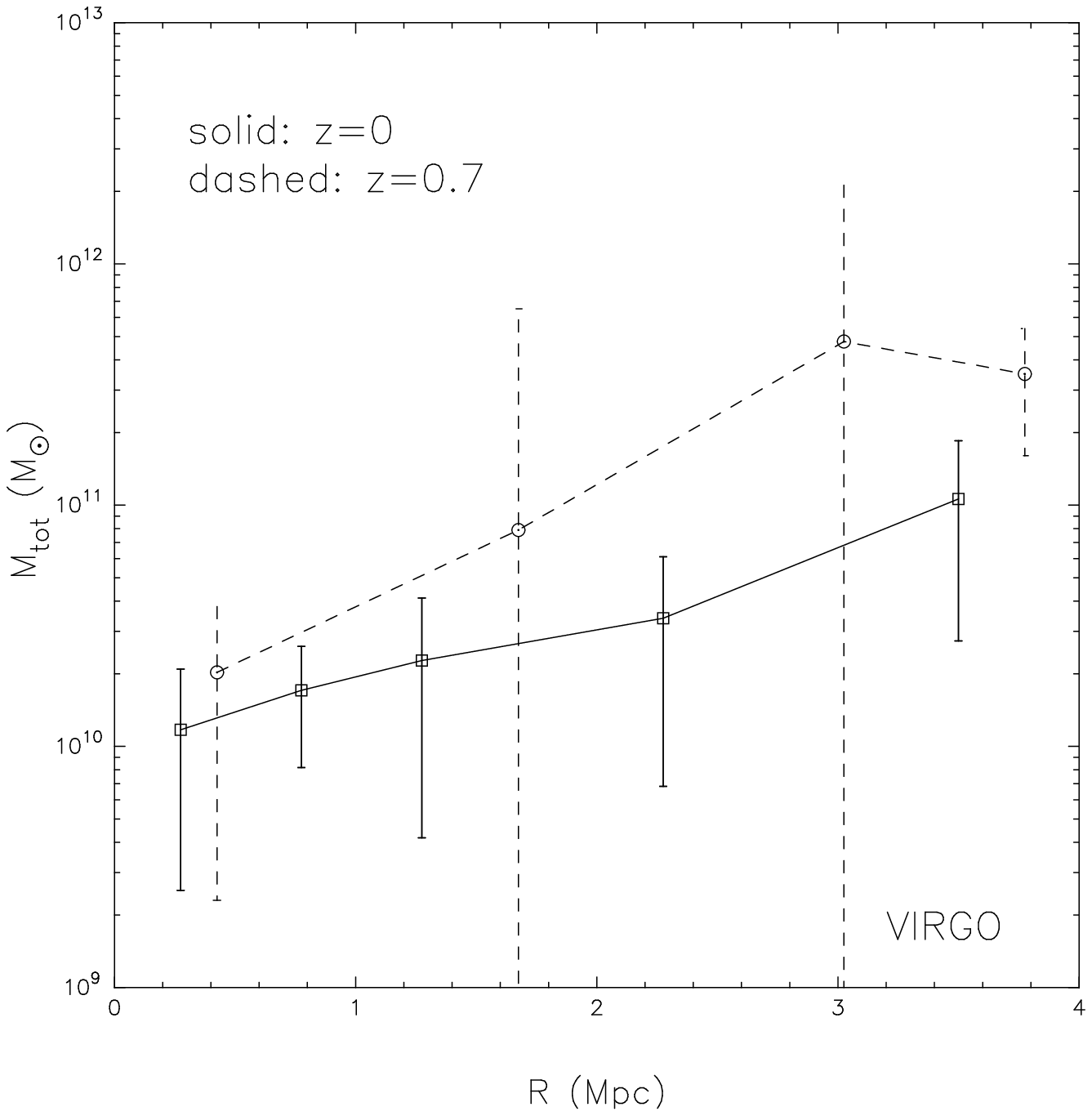}
\plotone{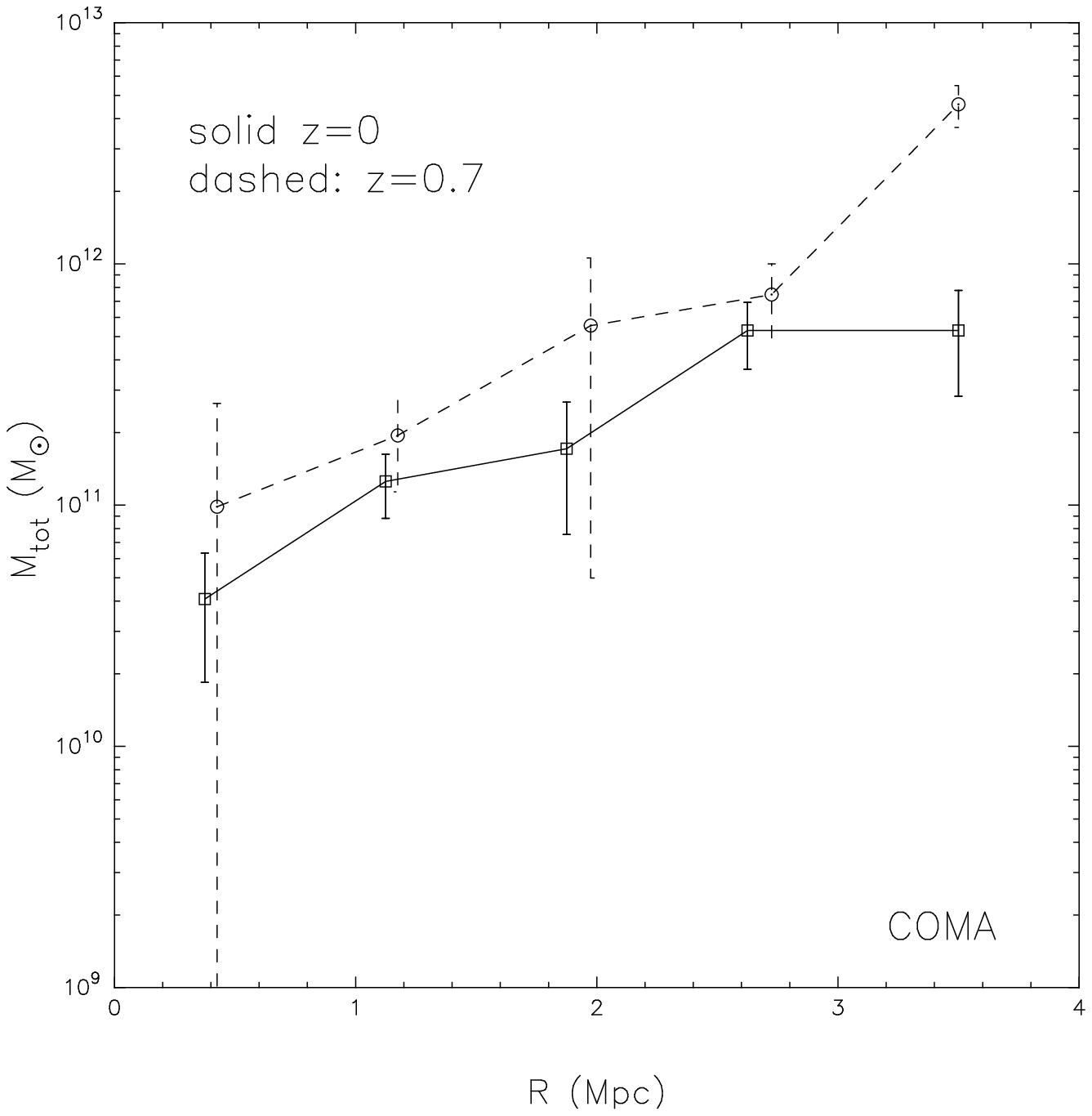}
\caption{Upper panels: median half-mass radius \rhalf (3D) and \Rhalf (2D) 
as a function of
projected cluster-centric distance, for $z$\,=\,0 and $z$\,=\,0.7; \virgo
(left) and \coma (right). For both clusters, the \rhalfRc and \RhalfRc trends are
clearly defined in \emph{both} redshift slices.
Also shown for comparison with
\virgo are the data point from G98 at $z$\,=\,0.
As we could have expected, we see that the 2D half mass radius \Rhalf is 
systematically smaller than the 3D half mass radius \rhalf.
Lower panels: median total mass as a
function of projected cluster-centric distance, for $z$\,=\,0 and
$z=0.7$; \virgo (left) and \coma (right). For both clusters at $z$\,=\,0,
we find a well defined (M$_{\mathrm{tot}}$, R) trend, though the
scatter is larger than in the \rhalfRc or \RhalfRc trends. At high redshift, the
scatter is larger and the (M$_{\mathrm{tot}}$, R) trend becomes marginal.}
\label{rhalf_Mtot_Rc}
\end{figure*}

The extent and efficiency of tidal stripping depends primarily on the
details of the orbital parameters of the subhalo. In this Section, we
study the evolution of the half mass radii (\rhalf, \Rhalf) and the
total mass \mtot as a function of the projected cluster-centric
distance R, for different redshift frames. In
Fig.~\ref{rhalf_Mtot_Rc}, we plot \rhalf and \Rhalf as a function of
\Rc both for \virgo and \coma, for the two redshift frames we
consider, i.e. $z$\,=\,0 and $z$\,=\,0.7. We clearly see that both the
\rhalfRc and \RhalfRc trends are well defined, at \emph{any} redshift.
Closer to the center of the cluster the extent of the dark matter
halos of the cluster galaxies is smaller. Thus closer to the
center of the cluster, stronger tidal stripping is experienced by
galaxies.
The \rhalfRc and \RhalfRc trends are illustrated on Fig.~\ref{rhalf_Mtot_Rc}
for the two extreme redshift frames considered in the simulations,
$z$\,=\,0 and $z$\,=\,0.7.
This choice is motivated by clarity of the figure and to enhance the evolution
between $z$\,=\,0 and $z$\,=\,0.7.
These trends are inferred in all redshifts frames extracted from the
simulations ($z$\,=\,0, $z$\,=\,0.2, $z$\,=\,0.4 and $z$\,=\,0.7).
Note that if this trend is inferred both at $z$\,=\,0 and $z$\,=\,0.7, then
it must be present for intermediate redshifts as well.

We see that for any given \Rc, \rhalf and \Rhalf\, are
systematically smaller at lower redshift compared to the high redshift
frame: by $z$\,=\,0, the galaxies have had more passages through the
cluster center, have been stripped more and thus have a smaller
extent. As expected, we see that the 2D half mass radius \Rhalf is
systematically smaller than the 3D half mass radius \rhalf.

In Fig.~\ref{rhalf_Mtot_Rc}, we plot the total mass M$_{\mathrm{tot}}$
as a function of \Rc both for \virgo and \coma, for the two redshift
frames we consider.  At $z$\,=\,0, we infer a (M$_{\mathrm{tot}}$, R)
trend, but the scatter is larger than the one associated with the
\rhalfRc trend. This difference in trends can be understood as
follows: consider a galaxy of mass $m$ with a circular velocity $v_c$
located in the cluster whose mass is $M$, we can relate the scatter on
the tidal radius ($\delta r_t/r_t$) to the scatter on the galaxy mass
($\delta m/m$). The tidal radius of the galaxy is proportional to:
\begin{equation}
r_t \,\,\propto \,\, \left( \frac{m}{3\,M} \right)^{1/3}
\label{rtequ}
\end{equation}
(Eq.~\ref{rtequ1}) and the mass of the galaxy is given by:
\begin{equation}
m \,\, \propto \,\, r_t \,\, v_{c}^{2},
\label{mequ}
\end{equation}
thus $r_t \propto r_{t}^{1/3} \,\,\, v_{c}^{2/3}$,
so $r_t \propto v_c$ and $\delta r_t \propto \delta v_c$.\\
Differentiating Eq.~\ref{mequ}, we can write:
\begin{equation}
\delta m \, = \, \delta r_t\,v_{c}^{2} \,+\, r_t\,\,2\,v_c\,\delta v_c \, \propto \, 3\,\delta v_c\,v_{c}^{2}
\label{deltam}
\end{equation}
Combining Eq.~\ref{mequ} and \ref{deltam}, we get:
\begin{equation}
\frac{\delta m}{m} \, = \, \frac{3\,\delta v_c\,v_{c}^{2}}{r_t\,v_{c}^{2}} \, = \, 3\,\frac{\delta r_t}{r_t}
\end{equation}

Therefore finding a larger relative scatter in the
(M$_{\mathrm{tot}}$, R) trend than in the \rhalfRc trend is to be
expected.  Note that at $z$\,=\,0.7, \coma is not well relaxed and
some galaxies have not experienced the tidal field of the cluster
potential yet. As a consequence, the outer radial bin contains some
infalling galaxies which have values of \rhalf and total masses
typical of field galaxy values (i.e. \rhalf $>200\,$kpc and
M$_{\mathrm{tot}}>$ 10$^{12}$ M$_{\sun}$ as obtained from
galaxy-galaxy lensing and satellite dynamics analysis, see Section 1).

In Fig.~2, we also show the results from G98 at
$z$\,=\,0. G98 studied a simulated galaxy cluster with virial radius
1.95\,Mpc and virial mass 2.8\,$\times$\,10$^{14}$\,M$_{\sun}$. These
properties makes it very similar to our \virgo cluster. Thus we
compare their results (data points with triangle) to ours on \virgo,
for the redshift frame at $z$\,=\,0 (solid line). 
We find significant differences between our results and the one from G98.
Before discussing these differences, we want to caution that both
works are different and difficult to compare. Therefore, the disagreement
between these studies cannot be used to assess quantitatively the 
impact of the baryonic component in the tidal stripping process.
We also note that there could be significant differences in subhalo
finding algorithms.
Kepping that in mind, three statements
can be made from this comparison: data points from G98 have larger
error bars and are systematically higher than ours. Moreover, the
clear trend we infer appears marginal in the work by G98. The
difference in the error bars comes from the statistics: within 2\,Mpc,
G98 identified $\sim 200$ halos whereas we have $\sim650$ halos within
the same radius. The difference in the value of \rhalf can be partly
due to the difference in resolution: G98 use particles whose mass is
8.6\,$\times$\,10$^{8}$\,M$_{\sun}$ whereas we use particles whose
masses are 4.4\,$\times$\,10$^{7}$\,M$_{\sun}$ and
3.2\,$\times$\,10$^{8}$\,M$_{\sun}$ for the stellar and dark matter
respectively.  The mass of our individual particles is 3-20 times
smaller, thus we resolve smaller galaxies than G98: they adopt a
minimum of 32 particles to define a halo whose individual properties
are relevant, which translates into a minimum mass of
2.7\,$\times$\,10$^{10}$\,M$_{\sun}$.  In \virgo we have 477 galaxies
of total mass smaller than this threshold.  Therefore our sample is
more complete in the inclusion of lower galaxy masses compared to the
G98 sample. Since smaller galaxies tend to have smaller extents, this
could explain the shift between our results.

In order to properly investigate the role of the baryonic component in the tidal
stripping process,
we would need a devoted dark matter only simulation of \virgo and \coma.
Such simulations are underway, and will be presented and analysed in a
forthcoming publication, where we will be able to quantify the differences expected.

\section{Scaling of \rhalf and \mtot with luminosity \& total mass to light ratio}

Parametric strong lensing studies
\citep[\emph{e.g.}][]{stronglensing,tyson98,Priya1,mypaperIII}
do include the galaxies living in the core of the cluster in the modeling
since they can locally pertub strong lensing features.
In order to reduce the number of free parameters in lens modeling, the
standard approach consists of scaling the galaxy parameters with
luminosity as follows:
\begin{equation}
r_{1/2} \,\, \propto \,\, L^{\alpha} \,\,\,\,\,\, \& \,\,\,\,\,\, \mathrm{M}_{\mathrm{tot}} \,\, \propto \,\, L^{\delta}
\label{scalingrelations}
\end{equation}

\begin{figure*}[h!]
\epsscale{0.5}
\plotone{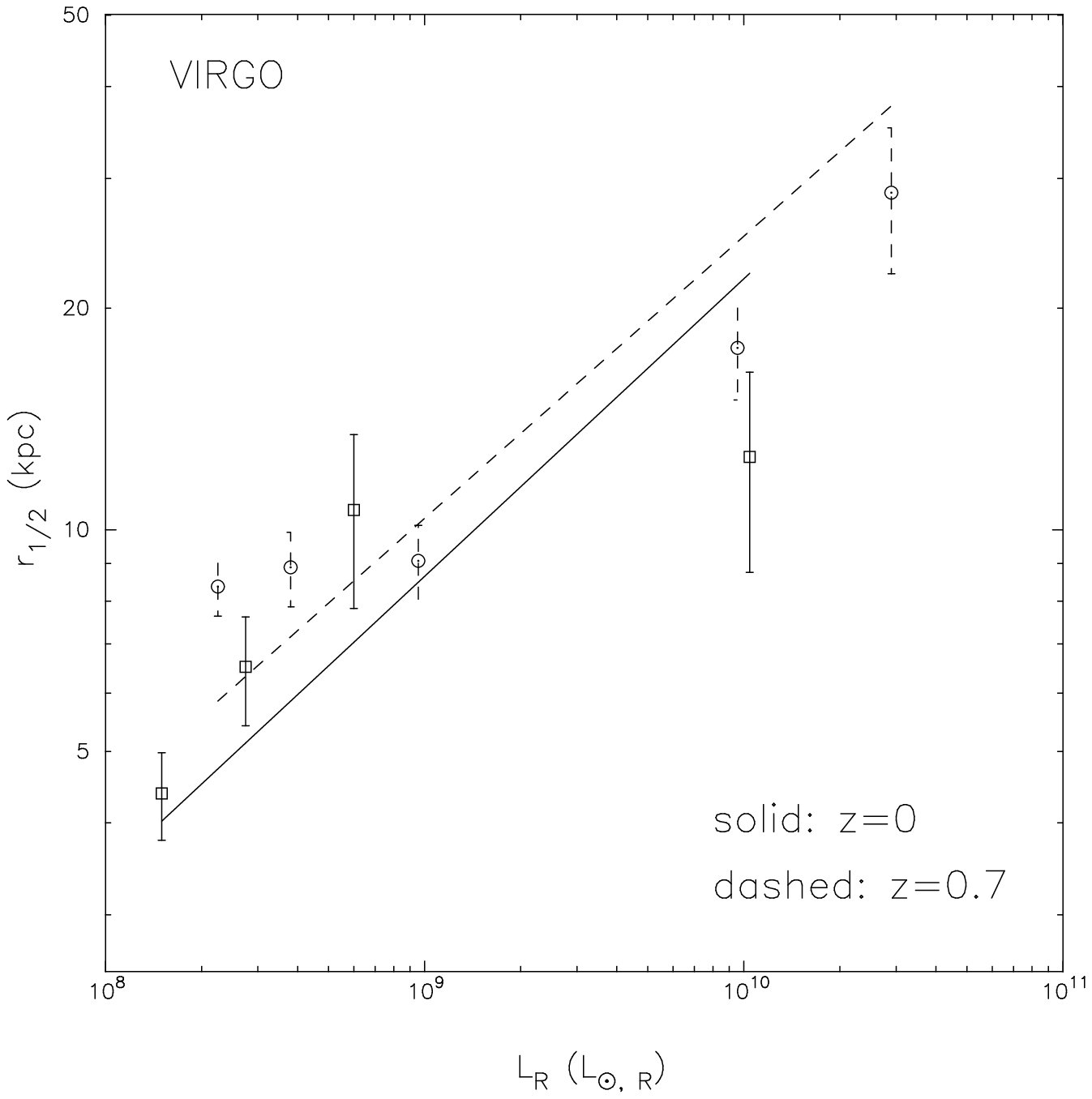}
\plotone{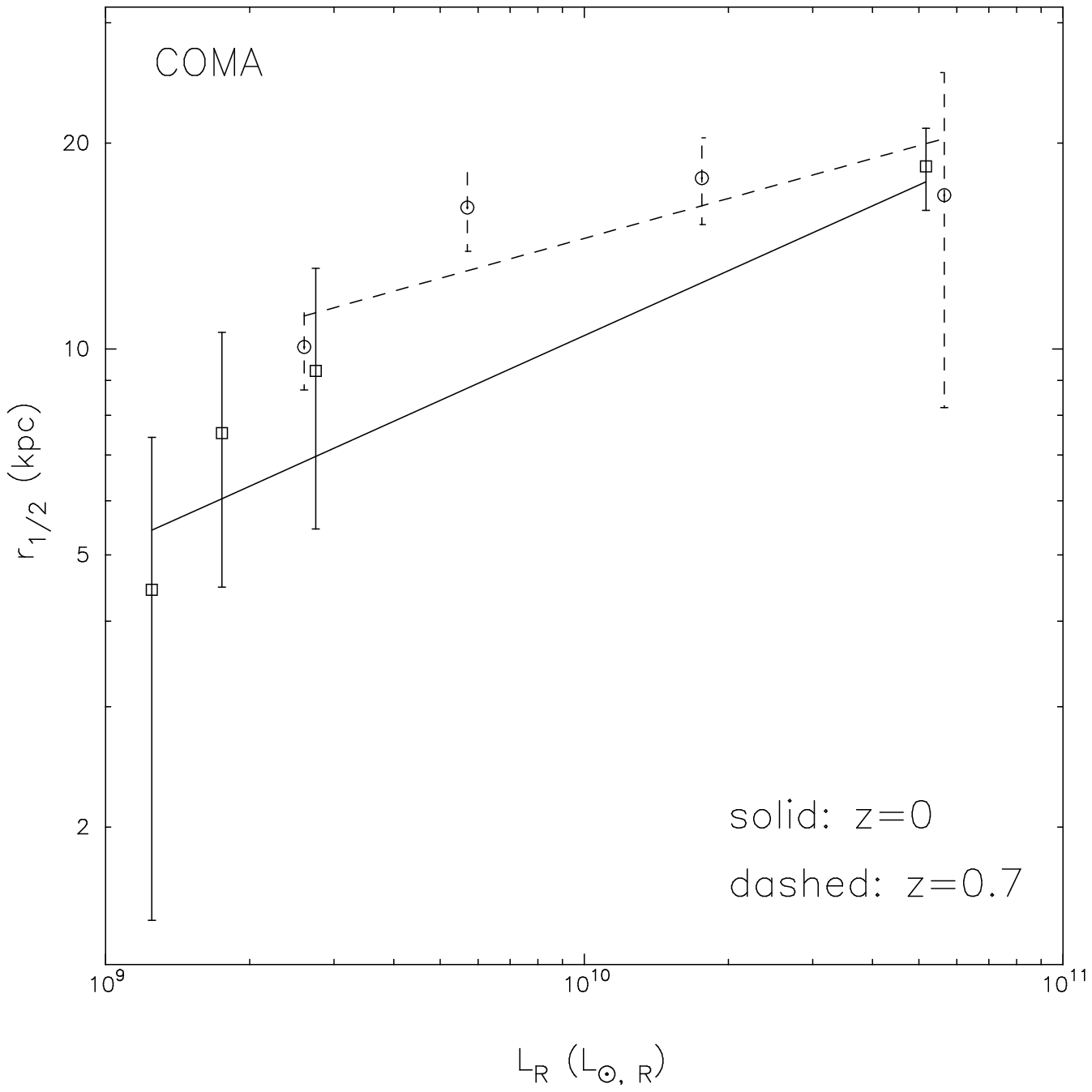}
\plotone{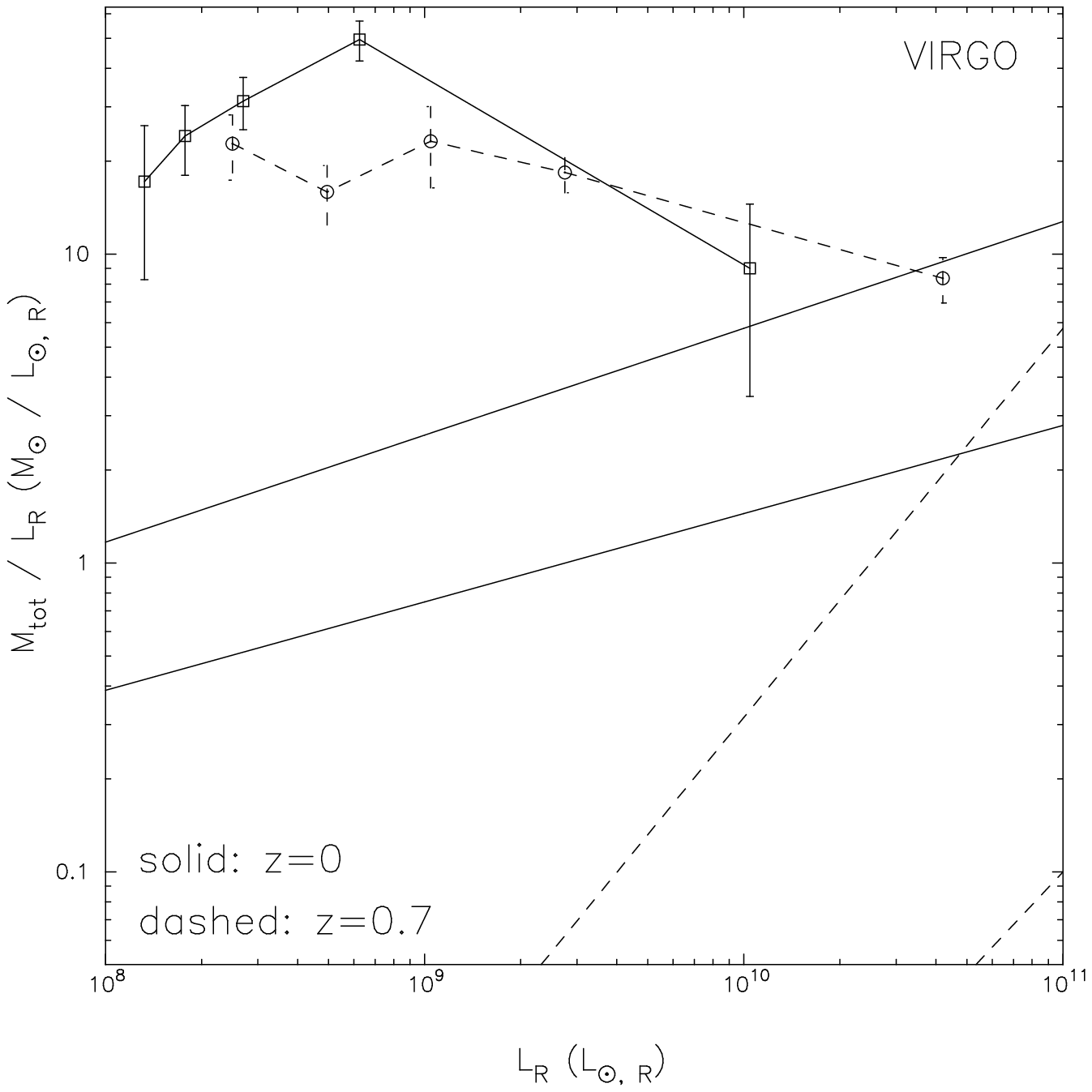}
\plotone{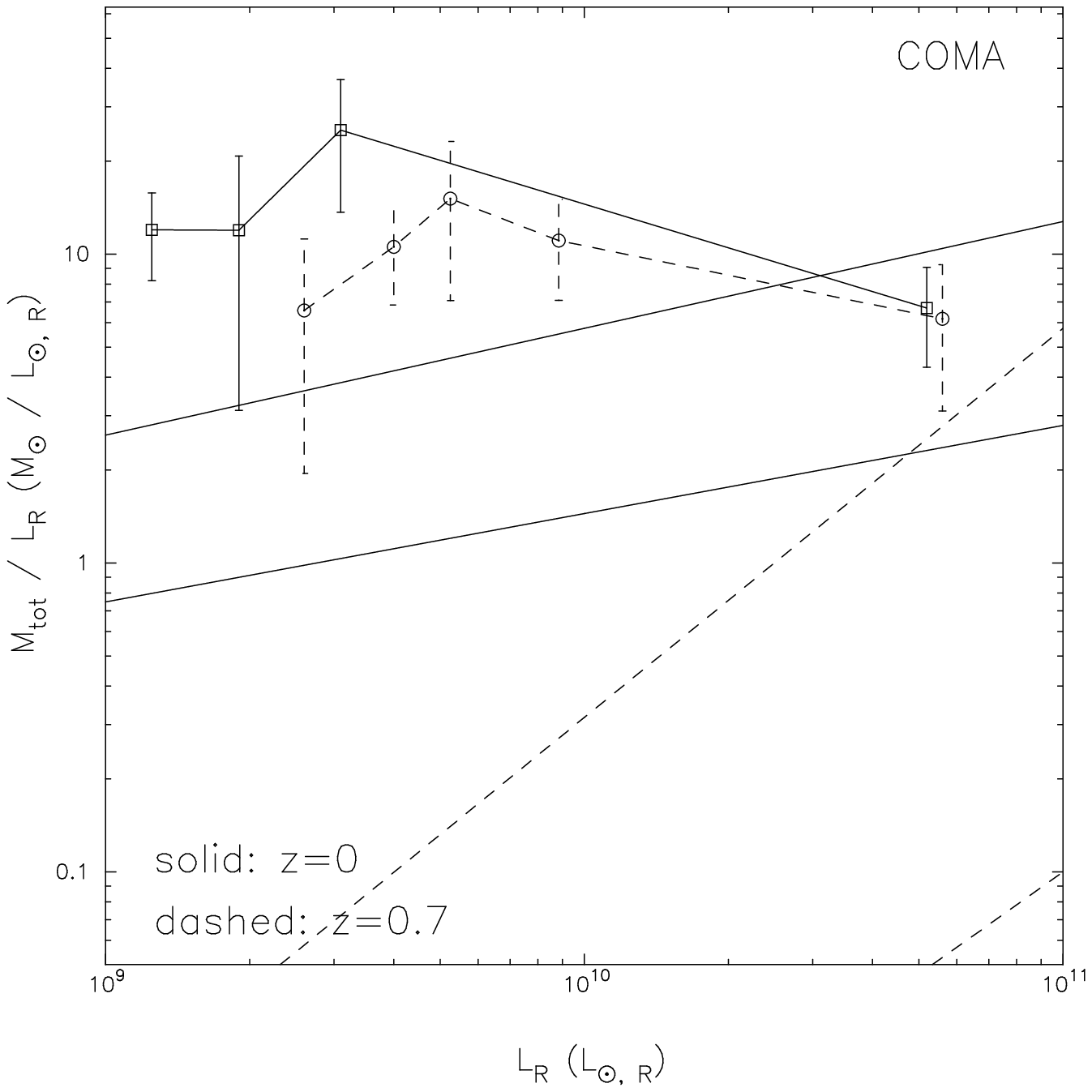}
\caption{
Upper panels: median half mass radius \rhalf as a function of luminosity 
for the central galaxies sample (\Rc $<$ 500\,kpc), 
for $z$\,=\,0 (solid squares) and $z$\,=\,0.7 (dashed circles); \virgo (left) and
\coma (right). Shown is the best fitting linear relation to the unbinned data
points (Table~\ref{lumfit}). We find that \rhalf increases
with luminosity, but has a larger scatter in the higher
redshift slice. Lower panels: median total mass
to light ratio as a function of luminosity for the central galaxies sample
for $z$\,=\,0 and $z$\,=\,0.7; \virgo (left) and \coma
(right). 
Overplotted lines correspond to the range allowed by the fundamental plane analysis
by \citet{jorgensen06}.
Note different axis scalings in the figure panels.
}
\label{Lum}
\end{figure*}

Galaxy-galaxy lensing studies also use similar scaling laws.  The
scaling usually used for the half mass radius in lensing studies is:
$\alpha=0.5$, which assumes that the mass to light ratio is constant
for all galaxies. At the present time strong lensing studies are not
in a position to discriminate between different scaling relations
\citep{aleksi,mcmc}.  In order to study the possible value of the
exponents defined by Eqn.~\ref{scalingrelations} that could be used in
strong lensing modeling, we want to examine the evolution of the half
mass radius \rhalf and the total mass \mtot with luminosity $L$, as a
function of redshift. We consider the galaxies located in the core of the cluster.
(i.e. the galaxies that satisfy \Rc $< 500$ kpc) since strong lensing
modeling deals with the inner part of a galaxy cluster (typically the
central 100$\arcsec$). In Fig.~\ref{Lum} (upper panels), we plot
\rhalf as a function of luminosity $L$ both for \virgo and \coma, at
$z$\,=\,0 and $z$\,=\,0.7. We fit a linear relation to the unbinned
data points for each redshift frame (i.e. $z=0, 0.2, 0.4, 0.7$) in
order to probe the exponent of the scaling relations defined by
Eqn.~\ref{scalingrelations}. The results are presented in
Table~\ref{lumfit}, and the best linear fit is plotted on
Fig.~\ref{Lum}. We also fit a linear relation to the unbinned data
points corresponding to the (M$_{\mathrm{tot}}$, $L$) trend
(Table~\ref{lumfit}). We find increasing (\rhalf, $L$) and
(M$_{\mathrm{tot}}$, $L$) trends at both redshifts, for the \coma and
\virgo central populations. Looking at the evolution of $\alpha$ and
$\delta$ with redshift, we do not infer any clear trend of evolution
of these values with redshift, thus we average the exponents from the
different redshift frames and we consider the averaged values in what follows.
Comparing \virgo and \coma, we see that the scatter is
smaller in the case of \virgo which is more relaxed than \coma (see
error bars on the linear fitting results in Table~\ref{lumfit}).

We caution that we do not claim to have derived fundamental scaling
relations here, but rather have fitted to our data sets.  As can be
seen in Fig.~\ref{Lum}, it is clear that the relations likely have a
more complicated form than a linear fit. 

These inferred scalings can
be compared to the fundamental plane derived for early-type galaxies
in clusters.  The values of $\delta$ derived in this work are in good
agreement with the fundamental plane analysis at $z=0$ for early-type
galaxies \citep[$\delta = 1.35 \pm 0.15$;][]{JFK96}.

On Fig.~\ref{Lum} we plot the total mass to light ratio \mtotL as a
function of luminosity. We find that \mtotL can reach values as high
as 55, with a mean value around $\sim$ 20-30.  It is interesting to
compare these total mass to light ratios to stellar mass to light
ratios (M/L) inferred observationally.  Recent studies
\citep{jorgensen06,jorgensen07} present a relation between the galaxy
masses and the mass to light ratios, for galaxy clusters in the local
Universe ($z$\,=\,0) and in the high redshift Universe ($z$\,=\,0.8 -
0.9).  The relations given by \citet{jorgensen06} were derived in the
\textsc{b} band.  We converted their \textsc{b} band luminosities into
the \textsc{r} band luminosities using a constant
(\textsc{b\,-\,r}\,=\,2) and inverted the relations in order to get a
relation between the stellar mass to light ratio and the \textsc{r}
band luminosity:
\begin{equation}
\log(\frac{\mathrm{M}}{\mathrm{L_R}}) = \gamma \log(\mathrm{L_R}) + \varsigma
\label{FPequ}
\end{equation}
for $z$\,=\,0, we find:
$$
\gamma=0.316\pm0.03 \,\,\,\,\, \& \,\,\,\,\, \varsigma=-2.7  
$$
and for $z$\,=\,0.8:
$$
\gamma=1.18\pm0.08\,\,\,\,\, \& \,\,\,\,\, \varsigma=-13.1
$$

We plot on Fig.~\ref{Lum} the lines corresponding to the range allowed
by these relations.  Obviously the total mass to light ratios \mtotL
derived in this work are systematically higher than stellar M/L's
inferred observationally.  This can be understood by the fact that we
consider the total mass, corresponding to both the baryonic and the
dark matter components, whereas the observational $M/L$ ratio is the
stellar mass to light ratio.

We did the same exercise for the whole galaxy population (i.e. for all
cluster-centric radii) to compare the (\rhalf, $L$) and
(M$_{\mathrm{tot}}$, $L$) trends inferred for the whole population to
the one inferred for the central population (\Rc $<$ 500\,kpc).  
When considering the
whole galaxy population, we find increasing (\rhalf, $L$) and
(M$_{\mathrm{tot}}$, $L$) trends at $z$\,=\,0, both for \coma and
\virgo. On the other hand, at $z$\,=\,0.7, these trends are hardly
defined and the scatter dominates, specially in the case of the
(M$_{\mathrm{tot}}$, $L$) trend.

This comparison suggests that in the core of the cluster at high
redshift, the galaxy population is already well relaxed and
constitutes a homogeneous galaxy population, whereas the galaxies at
larger distances from the cluster center have more complicated
dynamics. It appears that clusters have a 'well behaved relaxed
nucleus' at high redshift. With time this relaxed behavior propagates
itself outward into the outskirts of the cluster, resulting at
$z$\,=\,0 in a overall 'well behaved' galaxy population.

\begin{table*}
\caption{Deriving the scaling relations defined in Eq.~\ref{scalingrelations}}
\begin{center}
\begin{tabular}{cccccc}
\hline
\noalign{\smallskip}
Redshift & 0 & 0.2 & 0.4 & 0.7 & \textsc{mean}\\
\hline
\noalign{\smallskip}
\noalign{\smallskip}
\coma   & \begin{tabular}{c} $\alpha=0.315 \pm 0.111$\\ $\delta= 1.431\pm0.119$ \end{tabular}& \begin{tabular}{c} $\alpha= 0.165 \pm 0.136$\\ $\delta= 1.297 \pm 0.148$\end{tabular} &\begin{tabular}{c} $\alpha=0.274 \pm 0.090$\\ $\delta=1.189 \pm 0.098$ \end{tabular}& \begin{tabular}{c} $\alpha=0.194 \pm 0.059$\\ $\delta=0.983 \pm 0.077$ \end{tabular} & \begin{tabular}{c} $\alpha=0.237 \pm 0.045$\\  $\delta=1.22 \pm 0.055$ \end{tabular} \\
\noalign{\smallskip}
\hline
\noalign{\smallskip}
\virgo   & \begin{tabular}{c} $\alpha=0.404\pm0.050 $\\ $\delta=1.303\pm0.072$ \end{tabular}& \begin{tabular}{c} $\alpha=0.213\pm0.036$\\ $\delta=0.905\pm0.068$ \end{tabular}& \begin{tabular}{c} $\alpha=0.275\pm0.036$\\ $\delta=0.963\pm0.064$ \end{tabular}& \begin{tabular}{c} $\alpha=0.382\pm0.031$\\ $\delta=1.186\pm0.053$ \end{tabular} & \begin{tabular}{c} $\alpha=0.318 \pm 0.02$\\  $\delta=1.09 \pm 0.03$ \end{tabular} \\
\noalign{\smallskip}
\hline
\end{tabular}
\label{lumfit}
\end{center}
\end{table*}
\vspace{1cm}

\section{Distribution of Stars and Dark Matter in Individual Galaxies}

In order to get insight on how tidal stripping modifies the different
galaxy mass components, we study the ratio of the total mass to the
stellar mass and its evolution with redshift and cluster-centric
distance. First, we verify that there is no dependence of the stellar mass with
\Rc. We find that the scatter in the stellar mass of the galaxy
sample is large, and there is no discernable trend with \Rc at any redshift,
suggesting that the stellar component, on average, remains roughly
unaffected by tides.

\begin{figure*}[h!]
\epsscale{0.5}
\plotone{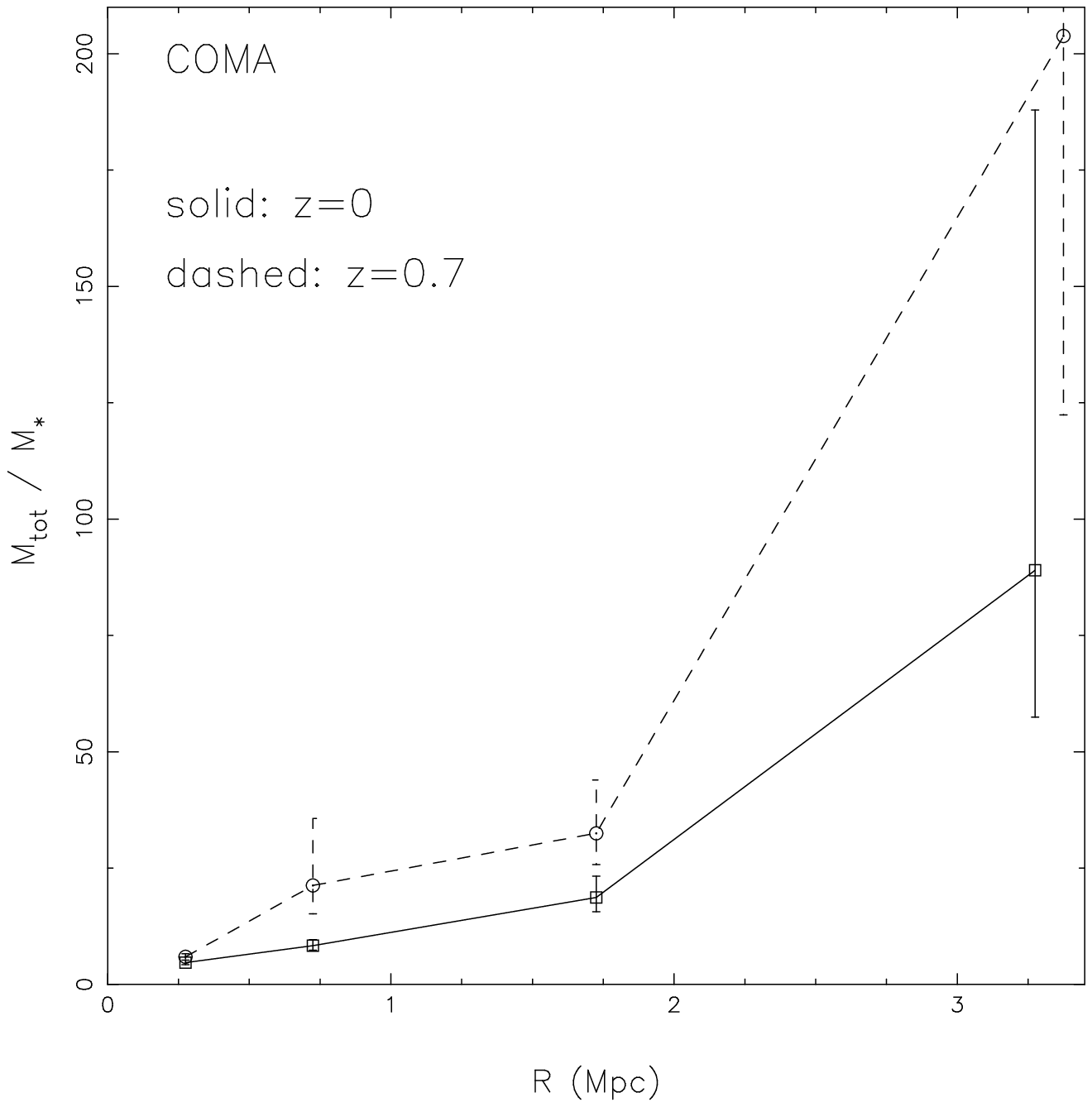}
\plotone{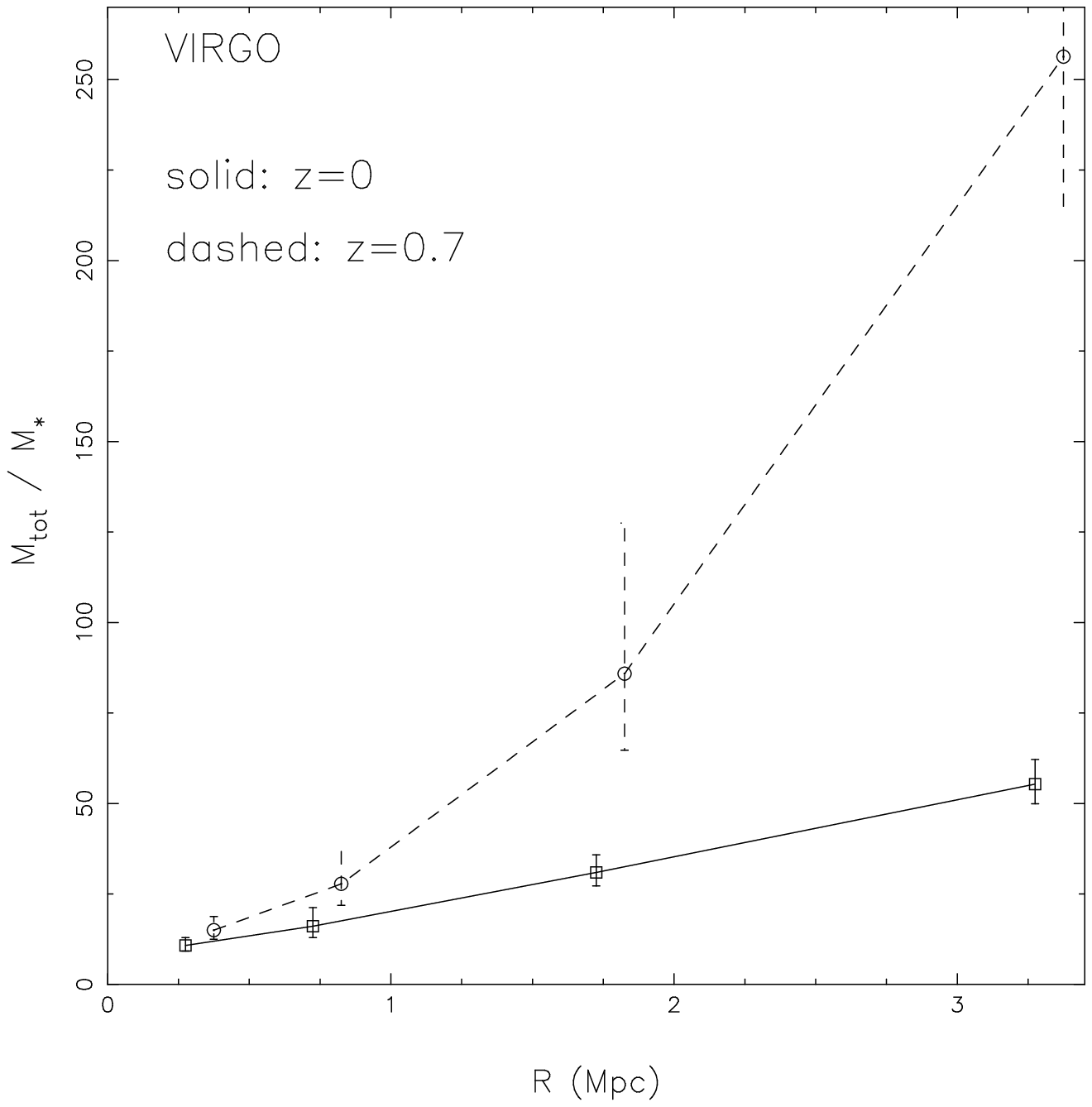}
\caption{Median ratio of the total mass to the stellar mass
of galaxies as a function of projected cluster-centric distance \Rc
at $z$\,=\,0 and $z$\,=\,0.7. Left: \coma; right: \virgo.}
\label{Mstar_Mtot_Rc}
\end{figure*}

Then, we study the ratio of the total to the stellar mass,
${\mathrm{M}_{\mathrm{tot}}}/{\mathrm{M}_{*}}$ and its evolution with
both cluster-centric distance and redshift. Fig.~\ref{Mstar_Mtot_Rc}
shows the result for $z$\,=\,0 and $z$\,=\,0.7. The ratio of the total to the stellar mass
clearly increases with cluster-centric
distance. Since tidal forces remove subhalo mass from the outside in
\citep{diemand07}, this result suggests that the extent of the dark
matter subhalo (quantified by \rhalf) is larger than the extent of the
stellar component (quantified by \rhalfstar, the half light radius) 
even in the very central
part of the cluster. To investigate this further, we compare the
extent of the stellar component \rhalfstar with \rhalf. We find
that \rhalfstar does not depend on the cluster-centric distance \Rc
for any redshift, both for \coma and \virgo. In addition, for \virgo,
we find that \rhalfstar is compact $<$ 2\,kpc, whereas \rhalf $>$
8\,kpc (Fig.~\ref{rhalf_Mtot_Rc}, upper left panel, central bin).  For
\coma, we find that \rhalfstar $<$ 3\,kpc, whereas \rhalf $>$ 11\,kpc
(Fig.~\ref{rhalf_Mtot_Rc}, upper right panel, central bin).  This
shows that for even the most stripped halos in the central bin, \rhalf
is still $\sim$ 4 times larger than \rhalfstar, suggesting that the dissipational 
stellar component is more compact and not as affected by tides.

\section{Comparison with galaxy-galaxy lensing results}

In this Section, we compare the results from our
simulated galaxy population with the results derived from observational
galaxy-galaxy lensing studies in clusters.
We first need to be sure we compare things that are comparable.
The quantity we use from the simulation is the 3D half mass radius \rhalf.
Former galaxy-galaxy lensing analyses performed through clusters have used a 
parametric model to fit the lensing data, the so-called Pseudo Isothermal
Elliptical Mass Distribution \citep[\textsc{piemd}, see \emph{e.g.}][]{mypaperI}.
This profile is parametrized using two characteristic radii, the core radius and the
scale radius (respectively $r_{\rm{core}}$ and $r_{\rm{cut}}$ in former articles).
In galaxy-galaxy lensing analyses, since we cannot constrain it, 
we usually  fix the core radius
to an arbitrary small value, making this \textsc{piemd} profile equivalent to
the \citet{bbs} profile. 
As a result of the galaxy-galaxy lensing fit, we get some constraints on this
scale radius.
It is easy to show 
\citep{ardis2218} that the scale radius for the \textsc{piemd} 
profile without a core radius is equal to the radius within which the 3D mass
equals half of the total mass. Thus this parameter inferred from the galaxy-galaxy
lensing analysis can be reliably compared to \rhalf from the simulation.
It can also be shown \citep{ardis2218}
that for a \textsc{piemd}
profile without a core radius, the radius within which the 2D mass equals half
of the total mass is found to be at 3/4 of the scale radius.

Of course, lensing is sensitive to the 2D projected surface mass density, and it
may be confusing to state that we infer a scale radius related to the 3D properties
of the halo from a lensing analysis.
However, the 2D projected mass density we are sensitive to with lensing is a function
of a scale radius that turns out to be the radius within which the 3D mass
equals half of the total mass.

We perform the comparison 
only with \coma since it is a 6 keV cluster and therefore more similar to the
massive, lensing clusters studied observationally than \virgo. Note that if need be, it is
possible to rescale the \rhalf values found for the \coma galaxies to
the temperature of a cluster probed observationally using the scaling
relation: \rhalf $\propto$ $T^{-1/2}$.  The temperature of a cluster
characterized by a velocity dispersion $\sigma$ scales as: $T \propto
\sigma^2$, and its mass M at a cluster-centric distance
$r_{\mathrm{c}}$ scales as: M $\propto\, \sigma^2\, r_{\mathrm{c}}$.
From Eq.~\ref{rtequ} and \ref{mequ}, we can derive that the tidal
radius of a galaxy orbiting in the cluster with a circular velocity
$v_c$ scales as:
\begin{equation}
r_t \,\, \propto \,\, \frac{v_{c}^{2/3}\, r_{t}^{1/3}}{\sigma^{2/3} r_{\mathrm{c}}^{1/3}}\, r_{\mathrm{c}}
\end{equation}
so $r_t\, \propto v_c\,r_{\mathrm{c}}\, / \, \sigma$, and finally
$r_t\,\propto T^{-1/2}$. This means that the more massive a galaxy
cluster is (thus the more dense the environment), the more severe the
tidal truncation.  For the purpose of the comparison with observed
clusters, all apparent magnitudes have been converted to the absolute
rest frame R band magnitude. We will apply the same
selection criteria (magnitude cut-offs) to the simulated galaxies in
order to compare the same luminosity galaxy population to the one
probed in galaxy-galaxy lensing studies. Finally, the galaxy-galaxy
lensing results have been rescaled to the median value of the
luminosity of the simulated galaxy population, using
the scaling laws defined in Eq.~\ref{scalingrelations} with $\alpha=0.5$
and $\delta=1.0$.

\begin{figure*}[h!]
\epsscale{0.5}
\plotone{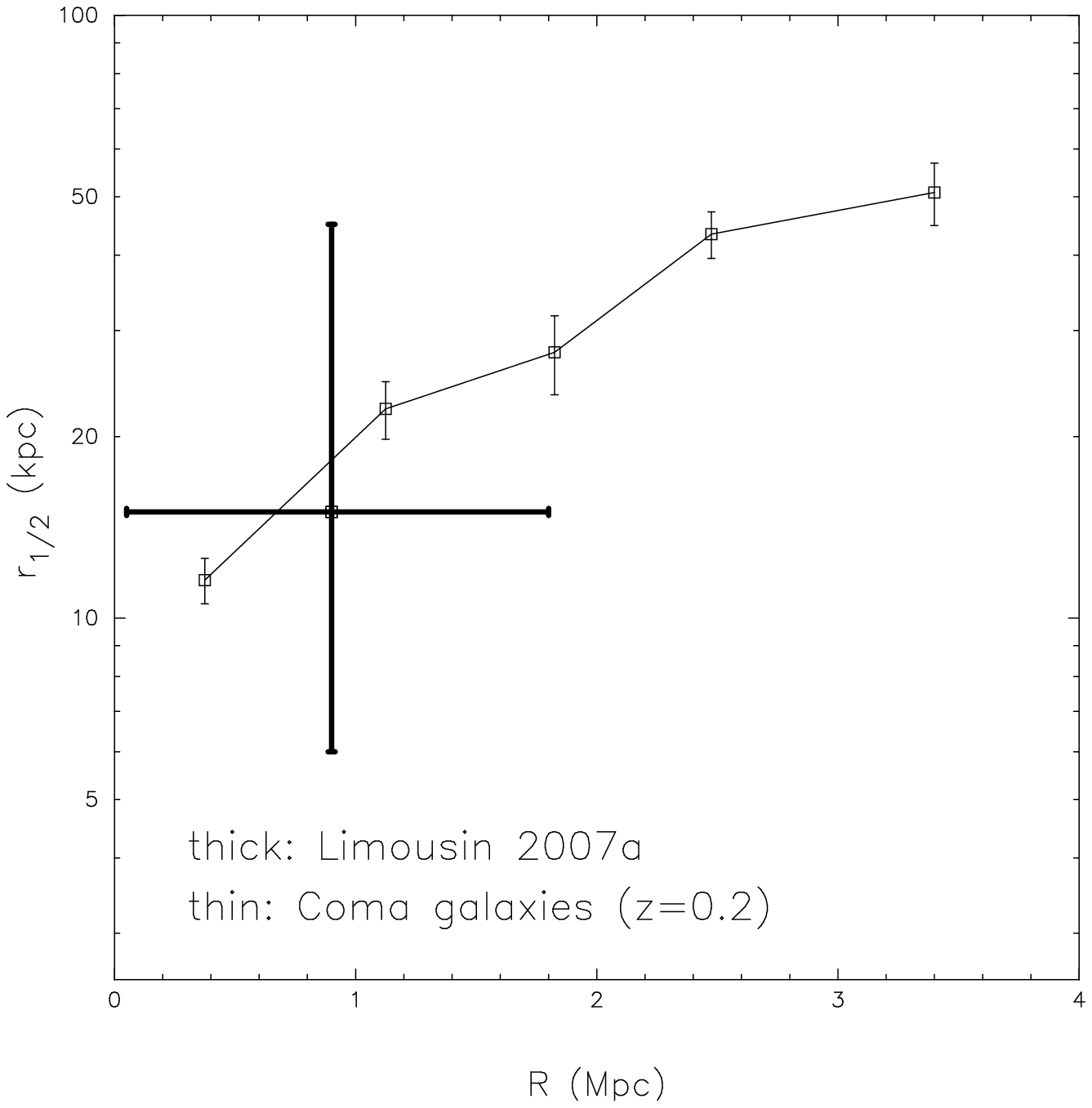}
\plotone{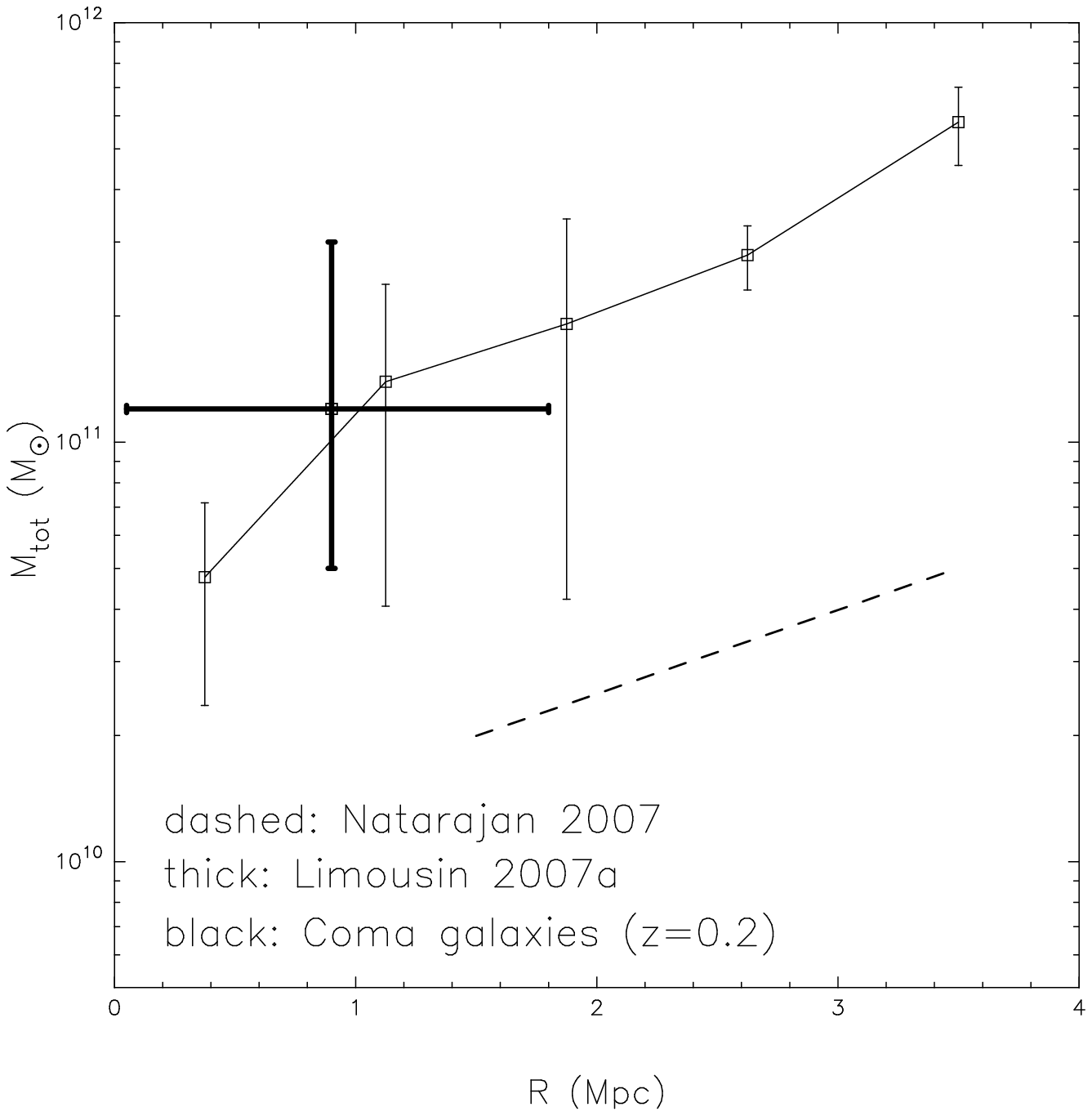}
\caption{Comparison of Limousin et~al.(2007a) and Natarajan
et~al.(2007) galaxy-galaxy lensing results with simulated \coma
galaxies. Left panel: \rhalf versus cluster-centric radius; Right
panel: total mass versus cluster-centric radius. To enable direct
comparison with the lensing data, the same magnitude cut-off has been
adopted in the simulations. The data point with error bars shown is
derived from galaxy-galaxy lensing results averaged over the subhalo
population ranging in cluster-centric radius from 0 to 1.8 Mpc for a
sample of massive lensing clusters. The slope indicated with a dashed
line shows the trend derived for Cl\,0024+16 from galaxy-galaxy
lensing techniques as well by Natarajan et al. (2007) using
space-based data. This trend is derived from mosaiced \textsc{hst wfpc-2} data
for Cl\,0024+16 out to 5 Mpc from the center. The slope corresponds to
the change in subhalo mass with cluster-centric radius for a typical
subhalo that hosts an early-type L$^*$ galaxy. Note that the
normalization cannot be compared directly due to the fact that the
central density of Cl\,0024+16 is significantly larger than that of
the simulated \coma studied here.}
\label{compartomine}
\end{figure*}

In particular, we compare the results presented in the previous
sections with the work by \citet{mypaperII}, which probes the cluster
population down to a magnitude of -17.5 for a sample of five galaxy
clusters at redshift around 0.2, with a mean \tsc{x}-ray temperature
of 8.5 keV.  As discussed in detail in \citet{mypaperII}, these
detections are in agreement with the similar studies performed by
Natarajan et~al. on observed clusters ranging from $z$\,=\,0.17 - 0.58
with space based \textsc{hst} data \citep{priya07}. Fig.~\ref{compartomine}
shows the comparison. We find qualitative agreement. However, due to
the weakness of the galaxy-galaxy lensing signal and the limitations
of ground based data, error bars on the estimated parameters are still
rather large. Also shown in Fig.~\ref{compartomine} is the slope
derived for the trend of the variation of typical subhalo mass (mass
of a subhalo that hosts an L$^*$ early-type galaxy) as a function of
cluster-centric radius for the massive lensing cluster
Cl\,0024+16 derived from space-based data. The data-set comprises of 
an HST mosaic that samples out to 5 Mpc from the cluster center and 
the constraints on the subhalo mass are derived applying galaxy-galaxy 
lensing methods in 3 radial bins. Strong and weak lensing constraints
are combined to derive the average properties of a typical dark matter
subhalo that hosts an L$^*$ early-type galaxy in each radial bin. The 
variation of the mass of the fiducial subhalo with cluster-centric
radius is well-fit by a linear relation. While the slope of this relation
can be directly compared to that derived from the simulated \coma cluster (as 
shown by the dashed line in Fig.~4) the normalization cannot be compared as 
the central density of Cl\,0024+16 is significantly higher.
We find that the trend derived from these space-based lensing observations
is in agreement with the results of the simulations. 

\section{Conclusions}

Using high resolution, hydrodynamical N-body simulations of two
fiducial galaxy clusters, one with parameters typical of \coma and the
other \textsc{virgo}, we study the tidal stripping process in detail.
These are the first simulations that include the baryonic component
that have been used to study this process, previous studies were
limited to dark matter only simulations.  We infer a strong trend
between the extent of cluster galaxy dark matter subhalos quantified
by the half mass radius \rhalf and the projected cluster-centric
distance \Rc. We show that the dark matter component is preferentially
stripped, whereas the stellar component is much less affected by tidal
forces. We infer a trend in these simulations that is much
stronger than the one inferred from the dark matter only cosmological
N-body simulations of \citet{ghigna98}. This could suggest that tidal
stripping is more efficient in the inner regions of clusters when the
effects of baryons are included.
However, we caution that comparing both simulations is not easy and therefore
we cannot assess reliably the impact of the baryonic component in the
tidal stripping process at this point. What is needed now is a devoted
dark matter only simulation of \virgo and \coma in order to compare
reliably the outputs of each study and investigate the expected influence
of the baryonic component on the tidal stripping process.

With the next generation of space telescopes, in particular with wide
field space based imagers, such as \textsc{dune}\footnote{www.dune-mission.net}
and \tsc{snap}\footnote{www.snap.lbl.gov}, we
will be able to probe this trend in a sample of galaxy clusters
spanning different range of masses and dynamical states. The limiting
aspect of these kind of studies are the relatively small mass of the
dark matter subhalos associated with cluster galaxies (leading to a
small modulation in the overall tangential shear field making
detection challenging), and therefore the large number of background
sources with reliable shape measurements needed for the lensing
analysis. A very promising technique will be the use of future radio
and millimeter-wave interferometers: accurate measurements of the
detailed dynamical structure of the background galaxies, in particular
rotating disks, should make it possible to probe 
the shear directly on each individual galaxy, thus allowing shear measurements
with much less galaxies than it is currently done \citep{blain02,morales}. 

From the numerical point of view, what is needed is a sample of galaxy
clusters that span a wide range in mass and dynamical states in a
large simulation box including baryonic physics, so that a robust
comparison between observations and theoretical expectations can then
be made. Such a quantitative comparison will be possible in the near
future and will provide further insights into the physics of cluster
assembly and the process of tidal stripping.

\section*{Acknowledgments}

The authors gratefully acknowledge the Dark Cosmology Centre at the
Niels Bohr Institute, University of Copenhagen. The Dark Cosmology
Centre is funded by the Danish National Research Foundation. We thank
Roser Pell\'o for providing us her code that allows to convert
magnitudes in different filters to absolute rest frame magnitudes and
for help in using it. We thank \'Ard\'is El\'iasd\'ottir, Gary Mamon, 
Jean-Paul Kneib, Jens Hjorth and Mark Wilkinson for useful discussions.
    
\bibliography{ms}

\end{document}